Delft University of Technology
Software Engineering Research Group
Technical Report Series

# Prioritizing Software Inspection Results using Static Profiling

Cathal Boogerd and Leon Moonen

Report TUD-SERG-2006-001

**T**UDelft

SERG

# Prioritizing Software Inspection Results using Static Profiling [*]

**Cathal Boogerd**

*Software Evolution Research Lab*
*Delft University of Technology*
*The Netherlands*
c.j.boogerd@ewi.tudelft.nl

**Leon Moonen**

*Software Evolution Research Lab*
*Delft University of Technology and CWI*
*The Netherlands*
Leon.Moonen@computer.org

**Abstract**

*Static software checking tools are useful as an additional automated software inspection step that can easily be integrated in the development cycle and assist in creating secure, reliable and high quality code. However, an often quoted disadvantage of these tools is that they generate an overly large number of warnings, including many false positives due to the approximate analysis techniques. This information overload effectively limits their usefulness.*

*In this paper we present ELAN, a technique that helps the user prioritize the information generated by a software inspection tool, based on a demand-driven computation of the likelihood that execution reaches the locations for which warnings are reported. This analysis is orthogonal to other prioritization techniques known from literature, such as severity levels and statistical analysis to reduce false positives. We evaluate feasibility of our technique using a number of case studies and assess the quality of our predictions by comparing them to actual values obtained by dynamic profiling.*

## 1. Introduction

*Software inspection* [14] is widely recognized as an effective technique to assess and improve software quality and reduce the number of defects [27, 19, 38, 26, 39]. Software inspection involves carefully examining the code, design, and documentation of software and checking them for aspects that are known to be potentially problematic based on past experience.

It is generally accepted that the cost of repairing a defect is much lower when that defect is found early in the development cycle. One of the advantages of software inspection is that the software can be analyzed even *before* it is tested. Therefore, potential problems are identified and can be solved early, when it is still cheap to fix them.

In this paper, we focus on tools that perform automatic code inspection. Such tools allow early (and repeated) detection of defects and anomalies which helps to ensure software quality, security and reliability. Most defect detection techniques are built around static analysis of the code. In its simplest form, this can be the warnings generated by a compiler set to its pedantic mode. In addition, various dedicated static program analysis tools are available that assist in defect detection and writing reliable and secure code. A well-known example is the C analyzer LINT [22]; others are discussed in the related work. These tools form a complementary step in the development cycle and have the ability to check for more sophisticated program properties than can be examined using a normal compiler; moreover, they can often be customized, and as such benefit from specific domain knowledge.

However, such static analyses come with a price: in the case that the algorithm cannot ascertain whether the source code at a given location obeys a desired property or not, it will make the safest approximation and issue a warning, regardless of the correctness. This conservative behavior can therefore lead to *false positives*, incorrectly signaling a problem with the code. Kremenek and Engler [24] observed that program analysis tools typically have false positive rates ranging between 30–100%. In addition, the increased scrutiny with which the code is examined can lead to an explosion in the *number* of warnings generated, especially when the tool is introduced later in the development process or during maintenance, when a significant code base already exists.

To cope with the large number of warnings, users resort to all kinds of (manual) filtering processes, often based on the perceived impact of the underlying fault. Even worse, our experiences indicate that the information overload often results in complete rejection of the tool, especially in cases where the first defects reported by the tool turn out to be false positives.

**Goal** In this paper, we aim at helping user of automated code inspection tools to deal with this information overload. Instead of focusing on improving a particular defect detection technique in order to reduce its false positives, we strive for a *generic prioritization approach* that can be applied to the results of any software inspection tool and assists the user in selecting the most relevant warnings.

To this end, we propose *ELAN*, a technique which orders inspection results based on a (demand-driven) static prediction of the *execution likelihood* of reported defects. We define execution likelihood to be the probability that a given program point will be executed at least once in an arbitrary program run. The rationale behind this approach is that the violating code needs to be *executed* in order to trigger the undesired behavior. As such, execution likelihood can be considered a *contextual measure* of severity, rather than the severity based on defect types that is usually reported by inspection tools.

[*] This work has been carried out as part of the Trader project under the responsibility of the Embedded Systems Institute. This project is partially supported by the Netherlands Ministry of Economic Affairs.

The ELAN algorithm is kept simple on purpose: scalability is an issue, as we want to be able to prioritize inspection results for large industrial systems. Analysis is done statically instead of dynamically because we want to prioritize inspection results already during development, when the system cannot yet be executed. In addition, the embedded nature of our industrial case makes dynamic profiling less suitable.

**Industrial Context**  The context of this work is the TRADER project in cooperation with Philips Semiconductors, in which we investigate and develop methods and tools for ensuring reliability of consumer electronics devices.

Modern consumer electronics such as MP3-players, mobile phones, televisions, and audio/video media centers, increasingly rely on embedded software for their operation. In the past, functionality of such devices was mostly implemented in hardware, but nowadays the features of these devices are made easily extensible and adaptable by means of software. As a consequence, the amount of software embedded in consumer electronics has grown tremendously.

For example, a modern television contains several million lines of C code and this amount is growing rapidly with new functionality, such as electronic program guides, increased connectivity with other devices, audio and video processing and enhancements, and support for various video encoding formats. During the development process, this code is routinely inspected using QA-C, one of the leading commercial software inspection tools currently on the market. Nevertheless, the developers have experienced problems handling the information overload mentioned earlier which motivated the research described in this paper.

**Overview**  The remainder of this paper is organized as follows: section 2 discusses related work, followed by an description of our approach in section 3. The ELAN algorithm is presented in section 4. Section 5 discusses a number of experiments and case studies conducted, and in section 6 the results and approach is evaluated. We conclude with an overview of contributions and future work in section 7.

## 2. Related Work

**Automatic Code Inspection**  There are a number of tools that perform some sort of automatic code inspection. The most well-known is probably the C analyzer Lint [22] that checks for type violations, portability problems and other anomalies such as flawed pointer arithmetic, memory (de)allocation, `null` references, and array bounds errors. LClint and splint extend the Lint approach with annotations added by the programmer to enable stronger analyses [12, 13]. Various tools specialize in checking security vulnerabilities. The techniques used range from lightweight lexical analysis [32, 37, 30] to advanced and computationally expensive type analysis [21, 17], constraint checking [34] and model checking [8]. Some techniques deliberately trade formal soundness for performance in order to scale to the analysis of larger systems [11, 6, 16] whereas others focus on proving some specific properties based on more formal verification techniques [3, 7, 9].

Several commercial offerings are available for conducting automated automatic code inspection tasks. Examples include QA-C,[1] K7,[2] CodeSonar,[3] and Prevent.[4] The latter was built upon the MECA/Metal research conducted by Engler et al. [41, 11]. Reasoning[5] provides a defect analysis *service* that identifies the location of potential crash-causing and data-corrupting errors. Besides providing a detailed description of defects found, they report on *defect metrics* by measuring a system's defect density and its relation to industry norms.

**Ordering Inspection Results**  The classic approach most automated code inspection tools use for prioritizing and filtering results is to classify the results based on *severity levels*. Such levels are (statically) associated with the *type* of defects detected; they are oblivious of the actual code that is being analyzed and of the location or frequency of a given defect. Therefore, the ordering and filtering that can be achieved using this technique is rather crude. Our approach is based on the idea that this can be refined by taking into account certain properties of the identified defect with respect to the complete source code that was analyzed.

A technique that is more closely related to our approach, is the z-ranking technique by Kremenek and Engler [24]. They share our goals of prioritizing and filtering warnings based on their properties with respect to analyzed code but do so based on the frequency of defects in the results. Their approach aims to determine the likelihood that a given warning is a false positive. It is based on the idea that, typically, the density of defects in source code is low. Thus, when checking source code for a certain problem, there should be a large number of locations where that check is not triggered, and relatively few locations where it is triggered. Conversely, if a check results in many triggered locations and few non-triggered ones, these triggers are more likely to be false positives. This notion is exploited by keeping track of success and failure frequencies, and calculating a numeric score by means of a statistical analysis. The warning reports can then be sorted accordingly.

Besides severity levels and z-ranking, we are not aware of any other work that deals with ordering inspection results.

**Static Profiling**  Static profiling is used in a number of compiler optimizations or worst-case execution time (WCET) analyses. By analyzing program structure, a prediction is made as to which portions of the program will be most frequently visited during execution. Since this heavily depends upon branching behavior, some means of branch prediction is needed. This can range from simple and computationally cheap heuristics to expensive data flow based analyses such as constant propagation [23, 28, 4] or symbolic range propagation [31, 5, 25]. Although there has been a lot of research on branch prediction, there are only a few approaches that take this a step further and actually compute a complete static profile [40, 35]. For branch prediction, these use heuristics similar to the ones we employ. The difference with respect to our work is that they try to estimate the, computationally expensive, execution *frequency*, whereas we compute execution *likelihood* which is less involved. Moreover, in contrast with these approaches, we do not perform computations for the complete program, but *demand-driven*: only for the locations associated with the warning reports.

---

[1] `www.programmingresearch.com`
[2] `www.klocwork.com`  [3] `www.grammatech.com`
[4] `www.coverity.com`  [5] `www.reasoning.com`

**Testability** Voas et al. [33] define *software testability* as "the probability that a piece of software will fail on its next execution during testing if the software includes a fault". They present a technique, dubbed *sensitivity analysis*, that analyses execution traces obtained by instrumentation and calculates three probabilities for every location in the program. Together they give an indication of the likelihood that a possible fault in that location will be exposed during testing. The first of these three, *execution probability*, is similar to our notion of execution likelihood, the chance that a certain location is executed. The other two are the *infection probability*, i.e. the probability that the fault will corrupt the data state of the program, and the *propagation probability*, the likelihood that the corrupted data will propagate to output and as such be observable.

Although the concepts involved are very similar to our own, the application and analysis method differ greatly: a location that is unlikely to produce observable changes if it would contain an error should be emphasized during testing, whereas we would consider that location to be one of low priority in our list of results. In addition, Voas approximates these probabilities based on dynamic info whereas we try to make predictions purely statically. Finally, infection- and propagation probability apply to locations that contain faulty code, while the inspection results we are dealing with may also reveal security vulnerabilities and coding standard violations that do not suit these two concepts.

## 3. Approach

The approach we propose for prioritizing code inspection results is based on the workflow depicted in figure 1. The process consists of the following steps (starting at top-left node):

1. The source code is analyzed using some code inspection tool, which returns a set of inspection results.

2. The inspection results are normalized to the generic format that is used by our tools. The format is currently very simple and contains the location of the warning in terms of file and line number, and the warning description. We include such a normalization step to achieve independence of code inspection tools.

3. We create a graph (SDG) representation of the source code that is inspected. Nodes in the graph represent program locations and edges model control- and data flow.

4. For every warning generated by the inspection tool, the following steps are taken:

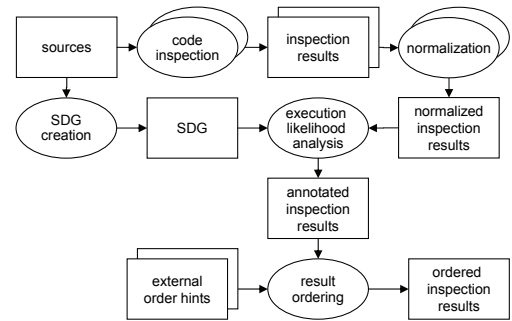

**Figure 1. Prioritization based on ELAN**

   (a) based on the reported source location, the analyzer searches the corresponding vertex in the graph.

   (b) it then proceeds to calculate the execution likelihood of this location/vertex based on analyzing the structure of the graph, and annotates the warning with this likelihood.

5. The inspection results are ordered by execution likelihood, possibly incorporating external hints such as severity levels or z-ranking results.

**System Dependence Graph** Central to our approach is the computation of the execution likelihood of a certain location in the program. In other words, we need to find all possible acyclic execution paths of the program that include our location of interest, and make predictions for the conditions (or branches) found along that path. To this end, we use the program's *System Dependence Graph* (SDG) [20], which is a generalization of the *Program Dependence Graph* (PDG).

In short, the PDG is a directed graph representing control- and data dependences within a single routine of a program (i.e. *intra-procedural*), and the SDG ties all the PDGs of a program together by modeling the *interprocedural* control- and data dependences. A PDG holds vertices for, amongst others, assignment statements, control predicates and call sites. In addition, there is a special vertex called *entry vertex*, modeling the start point of control for a function. In the remainder, we will use the terms vertex, program point and location interchangeably. The edges between vertices represent the control- and data dependences. Our approach currently does not consider information from dataflow analysis, so we limit our discussion to control dependences. The most relevant causes for such control dependencies are:

- there is a control dependence between a predicate vertex $v$ and a second vertex $w$ if the condition at $v$ determines whether execution reaches $w$;

- there is a control dependence between a function's entry point and its top-level statements and conditions;

- there is a control dependence between a call site and its corresponding function entry point.

Conditions in the code are represented by one or more *control points* in the SDG. Each *control point* corresponds to a 'simple' condition, e.g., `(a>1)||(b<3)` is represented by two control points (`a>1`) and (`b<3`) to ensure correct modeling of dependences within such short-circuited expressions.

Clearly, we can find all possible execution paths by simply traversing the SDG with respect to these control dependences. Moreover, they make for an efficient traversal, as we only need to visit those points that actually influence how control proceeds throughout the program, i.e. the control points and call sites. However, performing a traversal of the complete SDG for finding all paths to just a single point is not very efficient. To better guide this search, we base our traversals on *program slicing*.

**Slicing** The slice of a program $P$ with respect to a certain location $v$ and a variable $x$ is the set of statements in $P$ that may influence the value of variable $x$ at point $v$. Although we are not actually interested in dataflow information, this slice must necessarily include all execution paths to $v$, which is exactly what we are looking for. By restricting ourselves to control flow information, we can rephrase the definition as follows: the *control-slice* of $v$ in $P$ consists of all statements in $P$ that determine whether execution reaches $v$. Calculating the execution likelihood of $v$ is now reduced to traversing all paths within this slice. The next section presents an algorithm for performing such execution likelihood computations and discusses a number of optimizations to refine the results and to speed up computation.

## 4. Execution Likelihood Analysis

This section will introduce the algorithm calculating the execution likelihood for a single program point. Recall that we define the execution likelihood to be the probability that the given point will be executed at least once in an arbitrary program run. Given the SDG of a project, computation is based on a simple depth-first graph traversal, obtaining probabilities by predicting branch probabilities and combining all the paths found from the main entry point to our point of interest. For simplicity, we assume that the project contains a main function that serves as a starting point of execution. However, this is not a strict prerequisite, as we will see later on.

**Basic algorithm** To calculate an execution likelihood estimate $e_v$ for a vertex $v$ in a programs SDG $P$, we perform the following steps:

1. Let $B_v$ be the control-slice with respect to $v$. The result is a subgraph of $P$ that consists of the vertices that influence whether control reaches $v$.
2. Starting from the main entry point $v_s$, perform a depth-first search to $v$ within $B_v$, enumerating all the paths (sequences of vertices) to $v$. This is a recursive process; traversal ends at $v$, then the transition probabilities are propagated back to $v_s$, where the transition probability $p_{w,v}$ is the a posteriori probability that execution reaches $v$ via $w$. For any given vertex $w$ visited within the traversal, this is calculated in the following manner:

   (a) If $w$ is $v$, skip all steps, $p_{v,v}$ is 1.
   (b) For every control-dependence successor $s$ of $w$ in $B_v$, determine $p_{s,v}$
   (c) Determine the probability that control is transferred from $w$ to any of its successors $s$. We do this by first grouping the probabilities $p_{s,v}$ by the label of the edge needed to reach $s$ from $w$. E.g., when $w$ represents the condition of an if statement, we group probabilities of the true and false branches of $w$ together. For every group we determine the probability that at least one of the paths found is taken, we denote this set $S_w$.
   (d) If $w$ is not a control point, there will be just one element in $S_w$, and its probability is $p_{w,v}$.
   (e) If $w$ is a multiway branch (switch), all its cases are thought to be equally likely, and as such $p_{w,v}$ can be obtained by adding all probabilities in $S_w$ and dividing them by the number of cases.
   (f) If $w$ represents the condition of an if statement, we consider this a special case of the switch mentioned above: each of its branches is thought equally likely, so $p_{w,v}$ is obtained by adding both elements of $S_w$ and dividing them by 2.
   (g) If $w$ is a loop, it is assumed that the loop will be executed at least once. $S_w$ consists of one element representing the probability of the loop body, and $p_{w,v}$ is taken to be equal to this value.

3. When recursion returns at our starting point $v_s$, we have calculated the transition probability from $v_s$ to $v$, which is our desired execution likelihood $e_v$.

As stated earlier, the algorithm is designed for computation of execution likelihood in a project with a single starting point of execution. If we are dealing with a partially complete project, or we are in any other way interested in the execution likelihood with a different starting point, the approach can be easily modified to suit that purpose. Instead of using program slicing, we use a related operation called *program chopping*. The chop of a program $P$ with respect to a source element $s$ and a target element $t$ gives us all elements of $P$ that can transmit effects from $s$ to $t$. Notably, the chop of $P$ between its main entry point and any other point $v$ is simply the slice of $P$ with respect to $v$. Notice that if we let $B_v$ be the chop with respect to $v_s$ and $v$, we can take any starting point $v_s$ and end up with the desired conditional execution likelihood.

While traversing the graph, transition probabilities for the paths taken are cached. As such, when traversing towards $v$, for a given $w \in B_v$ we only need to calculate $p_{w,v}$ once. This approach necessarily only works within one traversal, i.e. when computing the execution likelihood of one point, because the prequel to some subpath may differ between traversals. However, when computing the likelihood for multiple locations within one program in a row, it is likely that at least part of the traversal results can be reused. For any point $v$ in a procedure $f$, we can split the transition probabilities into one from $v_s$ to the entry point $s_f$ of $f$, and the transition probability from $s_f$ to $v$. Effectively, this means that for any point in $f$ we only compute $p_{v_s,s_f}$ once.

An important contributor to performance is that our algorithm is *demand-driven*: it only computes execution likelihoods for locations of interest, instead of computing results for every location in the program. Given that the number locations with issues reported by an inspection tool will typically be much smaller than the total number of vertices in the graph, this is a sensible choice. Together with our deliberately simple heuristics, it forms the basis for a scalable approach.

**Refined branch prediction heuristics** To gain more insight into the performance/accuracy trade-off, the algorithm was extended with some of the branch prediction heuristics of Ball and Larus [2], used in the manner discussed by Wu and Larus [40]. They tested the heuristics empirically and used the observed accuracy as a prediction for the branch probability. For example, they observed that the value check heuristic predicts 'branch not taken' accurately 84% of the time. Therefore, when encountering a condition applicable to this heuristic, 16 and 84 are used for the 'true' and 'false' branch probabilities, respectively. Whenever more than one heuristic applies to a certain control point, the predictions are combined using the Dempster-Shafer theory of evidence [18], a generalization of Bayesian theory that that describes how several independent pieces of information regarding the same event can be combined into a single outcome.

The heuristics will replace the simple conventions used in steps (d) through (g) discussed above. The behavior with regard to multiway branches has not been changed, and in cases where none of the heuristics apply the same conventions are used as before. A brief discussion of the application of the different heuristics follows below, their associated branch prediction probabilities can be found in table 1. The table lists the probability a condition that satisfies the heuristic will evaluate to *true*. We should remark that these numbers are based on empirical research on different programs [2] than used in our experiments. However, Deitrich et al. [10] provide more insight into their effectiveness and applicability to other systems (and discuss some refinements specific to compilers).

The refined heuristics are:

*Loop branch heuristic:* This heuristic has been modified to apply to any loop control point. The idea is that loop branches are very likely to be taken, similar to what was used earlier. The value is used as multiplier for probability of the body.

*Pointer heuristic:* Applies to a condition with a comparison of a pointer against null, or a comparison of two pointers. The rationale behind this heuristic is that pointers are unlikely to be null, and unlikely to be equal to another pointer.

*Value check heuristic:* This applies to a condition containing a comparison of an integer for less than zero, less than or equal to zero. This heuristic is based on the observation that integers usually contain positive numbers.

*Loop exit heuristic:* This heuristic has been modified to apply to any control point within a loop that has a loop exit statement (i.e. break) as its direct control predecessor. It says that loop exits in the form of break statements are unlikely to be reached as they usually encode exceptional behavior.

*Return heuristic:* Applies to any condition having a return statement as its direct successor. This heuristic works because typically, conditional returns from functions are used to exit in case of unexpected behavior.

**Implementation** ELAN has been implemented as a plugin for Codesurfer,[6] a program analysis tool that can construct dependence graphs for C and C++ programs. As our approach is based on the SDG, the way in which this graph is constructed directly affects its outcome, especially in terms of accuracy. It should be noted, therefore, that the graphs both have missing dependences (false negatives) and dependences that are actually impossible (false positives). For example, control- or data dependences that occur when using setjmp/longjmp are not modeled. Another important issue is the accuracy of dependences in the face of pointers, think for example of modeling control dependences when using function pointers. To improve this accuracy, a flow insensitive and context insensitive points-to algorithm [29] is employed to derive safe information for every pointer in the program.

## 5. Experiments and Case Studies

This section reports on experiments designed to evaluate the accuracy and performance of our technique. Recall that the execution likelihood predictions are used in ranking different locations; we therefore compare rankings based on predictions with rankings based on measurements in actual program runs. Apart from this *ordering*, we also evaluate accuracy by checking correspondence of the actual prediction *values* with their measured counterparts. We analyze performance by timing the aforementioned experiments and relate the analysis speed to the size of the SDG involved. Finally, in sections 5.3 and 5.4 we present two case studies that were conducted to assess the approach in practice.

### 5.1. Correlating predictions with runtime

For this experiment we created a benchmark set of programs consisting mainly of simple open source command-line tools. Table 2 lists some source code properties for the different programs. Comment and whitespace lines were left out in

| Heuristic | Probability | Heuristic | Probability |
|---|---|---|---|
| Loop branch | 0.88 | Pointer | 0.40 |
| Value check | 0.16 | Loop exit | 0.20 |
| Return | 0.28 | | |

**Table 1. Heuristics and associated probabilities**

---

[6] www.grammatech.com

Boogerd, Moonen – Prioritizing Software Inspection Results using Static Profiling

| Project Name | KLoC | # vertices | # non-global | # CPoints | CC/fn | | | CPoint/LoC | | |
|---|---|---|---|---|---|---|---|---|---|---|
| | | | | | avg | stddev | max | avg | stddev | max |
| Antiword | 24 | 119391 | 35371 | 2787 | 8.0 | 11.0 | 596 | 0.13 | 0.10 | 0.77 |
| Chktex | 8 | 30422 | 10149 | 769 | 9.8 | 18.2 | 412 | 0.19 | 0.13 | 0.85 |
| Lame | 53 | 93812 | 39937 | 3673 | 7.9 | 25.9 | 904 | 0.12 | 0.19 | 3.25 |
| Link | 17 | 88766 | 33647 | 3009 | 5.6 | 9.4 | 358 | 0.16 | 0.15 | 1.0 |
| Uni2Ascii | 3 | 10368 | 5022 | 138 | 29.3 | 33.4 | 1004 | 0.23 | 0.21 | 0.91 |

**Table 2. Case study programs and their metrics**

the number of KLoC reported. The apparent discrepancy between the number of KLoC and the size of the SDG in vertices can be largely attributed to the use of global variables. In the SDG, global variables are simulated by adding them as a parameter to every function in the project, and every parameter introduces 2 extra vertices. To get a better feeling for the consequences, we have filtered out these generated global parameter vertices in the 'non-global' column. These results are already more in line with what we would expect from the size in terms of KLoC. Of course, the number of routines, parameters, and conditional statements are yet another influence. In the next seven columns, this is quantified by the total number of control points in the program, the cyclomatic complexity (per function), and the control points related to number of lines in the program (again per function).

The programs were picked such that it would be easy to construct 'typical usage' input sets, and automatically perform a large number of test runs. For every case, at least 20 different test runs were recorded. Every one of the programs was subjected to the following steps:

1. Build the project using Codesurfer. This involves the normal build and building the extra graph representations used by our technique.
2. Build the project using gcc's profiling options, in order to obtain profiling information after program execution.
3. Run the ELAN algorithm for every control-point vertex in the project. This will give a good indication of analysis behavior distributed throughout the program (since it approximates predictions for every basic block). Moreover, this step was performed twice: with and without branching heuristics.
4. Gather a small dataset representing typical usage for the project, and run the program using this dataset as input. For all the program locations specified in step 3, determine the percentage of runs in which it was visited at least once. This last step uses gcov, which post-processes the profile data gathered by gcc's instrumentation.
5. Create two sets of program locations, the first sorted by prediction, the second by actual usage, and compare them using Wall's unweighted matching method [36]. This will give us a *correlation score* for different sections of the two rankings.

To illustrate this correlation score, consider the following example: suppose we have obtained the two sorted lists of program locations, both having length $N$, and we want to know the score for the topmost $m$ locations. Let $k$ be the size of the intersection of the two lists of $m$ topmost locations. The correlation score then is $k/m$, where 1 denotes a perfect score, and the expected score for a random sorting will be $m/N$. In our experiments, scores were calculated for the topmost 1%, 2%, 5%, 10%, 20%, 40% and 80%. The correlation scores for the different programs can be found in table 3, where the first number in every cell is the score without the use of heuristics, and the second is obtained using heuristics.

When considering the ranking data in table 3, we will denote the upper $n$% of the ranking as *block n*. We are first and foremost interested in the correlation scores of the top-most blocks of locations. Typically, a randomized ranking would produce correlation scores of 0.01, 0.02 and 0.05 for these top blocks, which is easily outperformed by the ELAN technique. However, there is a problem in using these correlation scores for evaluating the accuracy of the ranking. Consider Antiword, for example: in our test series, 3.8% of the locations involved were always executed, which means that the two top-most blocks in our table will consist entirely of locations with measured value 1. There is no way to distinguish between these locations, and as such, no way to compare rankings. Even if predictions are perfect, locations in our predicted block 1 could be anywhere in block 2 of the other ranking. In practice this will be less of a problem since these values are used in conjunction with, e.g., the reported severity. Still, it shows that is important not to look at just the ordering, but also at the distribution of actual values.

In table 4, the average execution likelihood for all the locations in a certain block are shown; again, the values represent the results without heuristics and with heuristics, respectively. We took the ranking based on our *computed* likelihood and calculated the average *measured* execution likelihood. The values shown in table 4 cannot be compared amongst programs, as they depend on the runtime behavior of the program.

| Portion | Antiword | Chktex | Lame | Link | Uni2Ascii |
|---|---|---|---|---|---|
| Upper 1% | .45 - .45 | .25 - .50 | .14 - .14 | .22 - .22 | 1.0 - 1.0 |
| Upper 2% | .44 - .44 | .22 - .33 | .11 - .11 | .11 - .11 | 1.0 - 1.0 |
| Upper 5% | .61 - .61 | .30 - .26 | .25 - .24 | .30 - .29 | .67 - .67 |
| Upper 10% | .50 - .46 | .34 - .32 | .26 - .26 | .59 - .55 | .29 - .29 |
| Upper 20% | .42 - .45 | .24 - .26 | .34 - .34 | .57 - .57 | .67 - .67 |
| Upper 40% | .52 - .49 | .40 - .41 | .46 - .45 | .60 - .61 | .57 - .57 |
| Upper 80% | .78 - .79 | .76 - .76 | .83 - .83 | .83 - .83 | .79 - .79 |

**Table 3. Correlation scores**

| Portion | Antiword | Chktex | Lame | Link | Uni2Ascii |
|---|---|---|---|---|---|
| Upper 1% | .91 - 1.0 | 1.0 - 1.0 | .71 - .75 | .87 - .91 | 1.0 - 1.0 |
| Upper 2% | .73 - .89 | .89 - .89 | .35 - .47 | .89 - .94 | 1.0 - 1.0 |
| Upper 5% | .68 - .83 | .83 - .78 | .35 - .36 | .89 - .90 | 1.0 - 1.0 |
| Upper 10% | .65 - .72 | .72 - .72 | .40 - .43 | .89 - .90 | 1.0 - 1.0 |
| Upper 20% | .49 - .59 | .62 - .61 | .39 - .40 | .81 - .82 | .87 - .87 |
| Upper 40% | .40 - .40 | .41 - .41 | .30 - .29 | .70 - .72 | .57 - .57 |
| Upper 80% | .28 - .47 | .47 - .46 | .35 - .35 | .57 - .56 | .44 - .44 |

**Table 4. Average measured execution likelihood**

For example, consider the differences between values for Antiword, which has 3.8% of its locations always executed, and Chktex, which has 33.5% of its locations always executed. What does matter, however, is the distribution within one program: we expect the locations ranked higher to have a higher actual execution likelihood, and, with some exceptions, exactly this correlation can be observed here.

Finally, in a last experiment we investigated the accuracy of our predictions. Initial analysis of the correlation data showed that it was distributed according to an inverse bell curve. Therefore we focus our attention on investigating accuracy of values at the borders, i.e. values close to 1, or close to 0. In other words, can we trust predictions that a certain location will always be executed, or vice versa, never be executed? This question is relevant because the former locations will always end up at the top of our ranking, so we want to limit the number of false positives there. Similarly, the latter locations will be ranked at the bottom, therefore never inspected, and we do not wish to skip locations that might turn out to be important. For all the locations with associated prediction values close to 1 we checked whether they were really always executed. Likewise, for locations having predictions close to 0, we calculated the percentage of locations that were actually never executed. For example, the row for interval '> 0.98' holds the percentage of locations that have a prediction value higher than 0.98 *and* were always executed. Data for both the original variant and the one using heuristics, is shown in tables 5 and 6 in the familiar fashion.

## 5.2. Performance measurements

The benchmark set used in the previous section consists of programs of different size (cf. table 2), which helps to understand the scalability of the approach. Recall our algorithm, which computes a slice, traverses the subgraph obtained, and derives predictions for conditions. This signifies the importance of the size of the SDG, rather than the number of KLoC. This relationship is illustrated in figure 2. Timing measurements were taken for every experiment in section 5.1, and we calculated the average time taken per location. All measurements were performed on a laptop with an Intel Pentium Mobile 1.6Ghz and 512Mb memory running MS Windows XP Pro. For completeness, we have also included data for the industrial case described in the next section.

| Interval | Antiword | Chktex | Lame | Link | Uni2Ascii |
|---|---|---|---|---|---|
| 1 | 100 - 100 | 100 - 100 | 100 - 100 | 94 - 91 | 100 - 100 |
| > 0.99 | 100 - 100 | 92 - 93 | 100 - 100 | 92 - 93 | 100 - 100 |
| > 0.98 | 100 - 100 | 92 - 93 | 100 - 100 | 92 - 93 | 100 - 100 |
| > 0.95 | 100 - 100 | 92 - 93 | 100 - 100 | 92 - 94 | 100 - 100 |

**Table 5. Accuracy 'always executed' predictions**

| Interval | Antiword | Chktex | Lame | Link | Uni2Ascii |
|---|---|---|---|---|---|
| 0 | 100 - 100 | 100 - 100 | 100 - 100 | 100 - 100 | 100 - 100 |
| < 0.01 | 71 - 71 | 42 - 41 | 69 - 69 | 35 - 44 | 89 - 100 |
| < 0.02 | 69 - 69 | 45 - 42 | 68 - 69 | 34 - 41 | 79 - 89 |
| < 0.05 | 69 - 69 | 47 - 45 | 69 - 68 | 33 - 37 | 67 - 80 |

**Table 6. Accuracy 'never executed' predictions**

## 5.3. Industrial Case Study

This section discusses how our approach performs on the industrial source code from Philips that runs embedded in one of the televisions used as a guinea pig in the Trader project. The subsystem used in these experiments is the application that is used to drive the infrastructure layer of the television.

Unfortunately, the embedded nature of the software meant that we could not easily obtain dynamic profiling data. For example, the gcc profiling options used above add instrumentation that log execution data to disc, but the TV has no disc. As a consequence, we could not repeat the static versus dynamic profiling correlation experiments that we used earlier to benchmark our technique.

We intend to address this issue by extending the monitor that was concurrently developed by our colleagues [1] to collect dedicated dynamic profiling information and send it out via our test TV's communication port. However, that is left as future work. For now, we discuss a number of characteristics derived from Philips source code by our analysis and show that these results are comparable to the ones from the other case study, leading us to believe that the approach will work equally well on the Philips code.

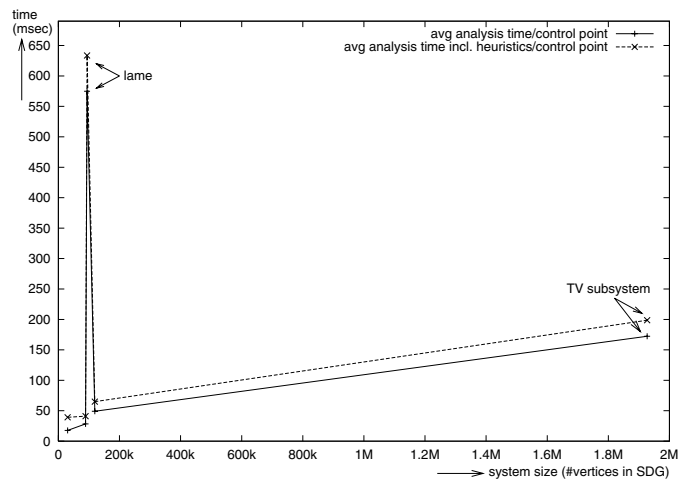

**Figure 2. Analysis time vs SDG size**

| Project Name | KLoC | # vertices | # non-global | # CPoints | CC/fn | | | CPoint/LoC | | |
|---|---|---|---|---|---|---|---|---|---|---|
| | | | | | avg | stddev | max | avg | stddev | max |
| TV subsystem | 67 | 1926980 | 119010 | 10079 | 9.6 | 19.1 | 909 | 0.33 | 0.51 | 10 |

**Table 7. Metrics for industrial case (TV subsystem)**

With respect to the metrics in table 7 we observe that again, we can see the impact of the use of global variables on the size of the SDG, and that the other characteristics are comparable to those in table 2. Figure 2 confirms this observation: the last point in the graph represents the television subsystem, and the approach seems to scale well to this graph size. It suggests that, in terms of performance, ranking locations in larger graphs is still feasible.

### 5.4. Inspection case study

This section reports on a small case study in ordering actual inspection results from SPLINT on two programs from our test set, Antiword and Lame. SPLINT uses lightweight static analysis to check a program for instances of well-known implementation flaws, and the checking can be extended by means of annotations. Since our two guinea pigs do not incorporate such annotations, SPLINT was run in *weak* mode, meant for "typical unannotated C code". The goal of this experiment is mainly to see whether ranking errors is feasible in terms of the time involved, and whether the accuracy remains at the level observed in our earlier correlation experiment (section 5.1).

In order to make this comparison, the same set of statistics are reported, they can be found in table 8. For both programs, the correlation scores and execution likelihoods are shown. In table 9, the analysis time per location in milliseconds is shown for the correlation case experiment discussed in section 5.1 (for reference), and for this inspection case. The last column shows the number of locations involved in the case study. The performances in the inspection case are illustrative for the effect of caching. This is apparent in the case of Lame, which benefits due to the larger number of locations, and the lack thereof in Antiword.

## 6. Evaluation

**Accuracy** When looking at the data of the accuracy experiments, perhaps most striking are the differences between tables 3 and 4. Even though locations with a higher execution likelihood in general seem to be ranked higher, the correlation scores resulting from the comparison with the ranking based on those measured values simply do not measure up. This trend is also visible in the rankings obtained in the case study. To understand this, we need to look at the locations that will end up high in the ranking: in our test set, the percentage of locations that were always executed ranged from 4% to 34%. This typically means that, even though we may propagate many of the 'interesting' locations towards the top of the list, we cannot distinguish between those in the topmost regions of that list. This explains the seeming discrepancy between correlation score and average execution likelihood.

A second interesting observation is that in all tests, prediction values close to 1 or 0 are good indicators of locations that will always be executed, or never be executed, respectively. The main problems for these predictions would be false positives or false negatives in the dependence graph used, but apparently this does not have a great impact. There is one notable anomaly, however: some locations in Link seem to have been mispredicted. Manual inspection revealed the culprits to be two top-level statements in the main function, which, judging by the flow of control and profile values reported for the surrounding statements, should always be executed, regardless of the programs input. Nevertheless, these lines are, seemingly incorrect, reported as never executed by gcov, resulting in the anomaly.

**Performance** There are two observations that we can make regarding performance: first of all, the approach seems to scale well to larger software systems, where the version that uses the refined branch prediction heuristics is only slightly outperformed by the simpler one. It goes to show that including more sophisticated analyses can still result in a feasible solution. Secondly, there is one program that does not conform to this observation: performance on Lame is significantly lower than on any of the others. Manual inspection revealed that the Lame frontend has a function that parses commandline arguments with a great number of short-circuited expressions. Because this occurs early on in the program, it affects many of the locations we are testing. Specifically, this means that computation of 25% of the locations involved requires traversal through this function, and 6.5% of the locations are within this function itself. The latter shows that our evalua-tion experiment is actually somewhat negatively biased since a relatively large number of locations were taken from a computationally expensive function. The data obtained from the case study in table 9 seems to corroborate this. However, this kind of biased distribution is unlikely when ordering actual inspection results. In addition, such large short-circuited expressions are atypical for the type of software analysed in the Trader project, leaving little reason for concern at this time.

| Portion | Antiword | | Lame | |
|---|---|---|---|---|
| | correlation | EL | correlation | EL |
| Upper 1% | 100 - 100 | 100 - 100 | 0 - 0 | 100 - 100 |
| Upper 2% | 100 - 100 | 100 - 100 | 11 - 11 | 100 - 100 |
| Upper 5% | 67 - 67 | 67 - 67 | 17 - 17 | 63 - 67 |
| Upper 10% | 50 - 50 | 50 - 50 | 44 - 41 | 45 - 42 |
| Upper 20% | 46 - 38 | 38 - 38 | 49 - 49 | 44 - 44 |
| Upper 40% | 54 - 50 | 20 - 20 | 69 - 70 | 26 - 27 |
| Upper 80% | 81 - 83 | 14 - 16 | 77 - 77 | 16 - 16 |

**Table 8. Inspection case: accuracy**

| Project | Correlation case | | Inspection case | | # Locations |
|---|---|---|---|---|---|
| | No heur. | Heur. | No heur. | Heur. | |
| Antiword | 52 | 69 | 360 | 531 | 74 |
| Lame | 594 | 655 | 313 | 329 | 542 |

**Table 9. Inspection case: performance (msec)**

**Orthogonality**  We remark that our approach is orthogonal to other prioritization and filtering techniques discussed in the related work. However, in combination with these approaches, ELAN can best be applied as final step because the filtering of the earlier stages in combination with our demand driven approach effectively reduces the amount of computations that will need to be done.

**Applicability**  In our test set, we use programs that are one-dimensional in their tasks, i.e. perform one kind of operation on a rather restricted form of input. This limits issues related to the creation of appropriate test inputs, and allows us to focus on evaluating the approach itself. It does mean, however, that we must devote some time to the question how to generalize these results to other kind of programs.

The ELAN approach is based on information implicit in the control structure of the program, and as for the heuristics, in the way humans tend to write programs. This information will always be present in any program. However, there may be parts of the control structure that are highly dependent on interaction or inputs. Fisher and Freudenberger observed that, in general, varying program input tends to influence which parts of the system will be executed, rather than influencing the behavior of individual branches [15]. This suggests that, typically, there are a number of highly data-dependent branches early on in the program, while the rest of the control structure is rather independent. For example, a command-line tool may have a default operation and some other modi of operation that are triggered by specifying certain command-line arguments. At some point in this program, there will be a switch-like control structure that calls the different operations depending on the command-line arguments specified. This control structure is important as it has a major impact on the rest of the program, and it is also the hardest to predict due to its external data dependence. However, this information (in terms of our example: which operation modi are most likely to be executed) is exactly the type of information possessed by domain experts such as the developers of the program. Therefore, the simple extension of our approach with a means to specify these additional (input) probabilities can further improve applicability to such situations.

## 7. Concluding Remarks

**Contributions**  We present a method for the prioritization of software inspection results based on statically computing the likelihood that program execution reaches locations for which issues are reported, i.e., we prioritize code inspection results using static profiling. We discuss a novel *demand-driven* algorithm for computing execution likelihood based on the system dependence graph and we present and evaluate a number of optimizations that further increase accuracy and performance.

We investigate the feasibility of the described approach using a number of case studies in which a prototype tool was developed and applied to several open source software systems. We investigate the correlation between our static predictions and actual execution data found by dynamic profiling and we report on the performance of our approach. This empirical validation shows that the approach is capable of producing a listing where program locations with a higher measured execution likelihood tend to be higher ranked. In addition, the approach scales well to larger systems.

Finally, we discuss experiments conducted on the Philips television software whose inspection motivated this research. We show that this code shares relevant characteristics with the code investigated in the case studies, leading us to believe that the approach will work equally well on the Philips code.

**Future Work**  Future work (that is actually already ongoing) includes deriving dynamic profiling data from the software embedded in a television and assessing the correlation between our execution predictions and the actual runtime data.

Together with Philips Semiconductors, we are preparing a larger case study, scheduled to start summer 2006, in which our technique will be evaluated during development and inspection of software to be embedded in a new digital TV.

Finally, we want to experiment with a number of ideas to further improve our approach by incorporating more advanced program analysis techniques, such as range propagation [25], that are basically aimed at enabling better estimations of the outcomes of conditions. Also, they can be used to compute execution frequencies, which will benefit the ranking by better distinguishing between locations with a likelihood of 1. However, since such analyses typically come with additional computational costs, we want to investigate if the improved accuracy actually warrants the expenses involved.

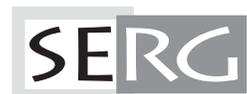